\documentclass[11pt]{article}
\usepackage{moriond,epsfig,amsmath}

\def\be{\begin{equation}}
\def\ee{\end{equation}}
\def\bea{\begin{eqnarray}}
\def\eea{\end{eqnarray}}

\def\tL{{\tilde L}}
\def\gappeq{\mathrel{\rlap {\raise.5ex\hbox{$>$}}{\lower.5ex\hbox{$\sim$}}}}
\def\lappeq{\mathrel{\rlap{\raise.5ex\hbox{$<$}}{\lower.5ex\hbox{$\sim$}}}}
\begin{document}
\vspace*{4cm}
\title{HIGHER-ORDER QCD CORRECTIONS FOR VECTOR BOSON PRODUCTION AT HADRON
COLLIDERS}
\author{ G.\ FERRERA}
\address{%Dipartimento di Fisica, 
Dipartimento di Fisica, Universit\`a di Firenze \& 
INFN Sez.\ di Firenze,\\
I-50019 Sesto Fiorentino, Florence, Italy}
\maketitle\abstracts{
We consider higher-order QCD corrections for vector boson production
at hadron colliders.
We present recent results on transverse-momentum 
resummation for $Z$ production. 
Moreover we 
show numerical results from a new fully exclusive next-to-next-to-leading 
order (NNLO) calculation.
  }
\section{Introduction}
The study of 
vector boson production in hadron collisions,
the well know Drell-Yan (DY) process, is nowadays extremely important. 
Owing to the large production rates and clean experimental signatures of 
$W$ and $Z$ productions, these processes are standard candles for calibration
purposes, they lead to precise determinations of vector boson masses and widths 
and provide important information on parton distribution functions (pdf).
It is therefore essential to have accurate theoretical predictions
for vector boson cross section and distributions.
This requires the computation of 
higher-order QCD radiative corrections for such processes.

In these proceedings we present two recent results on higher-order corrections
for the DY process:
a study of transverse-momentum ($q_T$) resummation for $Z$ production 
at the Tevatron~\cite{wI} 
and a fully exclusive calculation~\cite{wII}
based on the NNLO subtraction formalism of Ref.~\cite{hnnlo}.

\section{Transverse-momentum distribution: fixed-order and resummation}
%\clearpage
We consider the inclusive hard-scattering process
\begin{equation}
h_1(p_1) + h_2(p_2) \;\to\; V (M,q_T ) + X \;\to\; l_1+l_2 + X,   
\label{first}
\end{equation}
where $h_1$ and $h_2$ are the colliding hadrons with momenta
$p_1$ and $p_2$, $V$ is a vector boson
(which decays in the lepton pairs $l_1,l_2$)
with invariant mass $M$ and transverse-momentum $q_T$
and $X$ is an arbitrary and undetected final state. 

According to the QCD factorization theorem
the $q_T$ differential cross 
section $d\sigma^V/dq_T^2$ can be written as
\begin{equation}
%\label{fac}
\frac{d\sigma^V}{d{q_T^2}} (q_T,M,s) =
\sum_{a,b}\int_0^1dx_1\int_0^1dx_2 
\,f_{a/h_1}( x_1,\mu_F^2)\,f_{b/h_2}(x_2,\mu_F^2)
\,{\frac{d{\hat\sigma}^V_{ab}}{d{q_T^2}}} (q_T,M,\hat s;\alpha_S,\mu_R^2,\mu_F^2)
\nonumber
\end{equation}
where $f_{a/h}(x,\mu_F^2)$ are the parton densities 
 of the colliding hadrons 
at the factorization scale $\mu_F$, 
$d\hat\sigma^V_{ab}/d{q_T^2}$ are the perturbative QCD computable
partonic cross sections, 
$s$ ($\hat s = x_1 x_2 s$) is the hadronic (partonic) centre-of-mass  energy, 
and $\mu_R$ is the renormalization scale. 

\begin{figure}[h]
\begin{center}
\psfig{figure=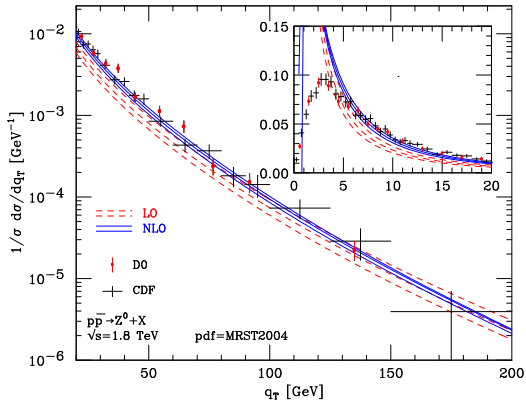,height=2.1in}
\caption{The $q_T$-spectrum of the Drell-Yan $e^+ e^-$ pairs produced in $p\bar p$ 
collisions at the Tevatron Run I. %~\cite{data}. 
Theoretical results are shown at LO %(red dashed lines)
and NLO %(blue solid lines) 
including the following scale 
variations: %
$m_Z/2 \leq \mu_F , \mu_R \leq  2\;m_Z$ , with the
constraint $1/2 \leq \mu_F /\mu_R \leq 2$.
\label{fo}}
\end{center}
\end{figure}

In the region where $q_T \sim  m_V$, $m_V$ being the mass of the 
vector boson ($m_V=m_W, m_Z$), 
the QCD perturbative
series is controlled by a small expansion parameter, 
$\alpha_S(m_V)$, and fixed-order calculations are
theoretically justified. In this region, 
the QCD radiative corrections are known up to next-to-leading order 
(NLO)~\cite{nlo}. 
From Fig.~\ref{fo} we see that the NLO result~\cite{wI} is in agreement with 
the experimental data~\cite{data} over a
wide region of transverse momenta 
($q_T \gappeq 20~{\rm GeV} $). 

In the small-$q_T$ region ($q_T\ll m_V$), 
the LO and NLO calculations do not
describe the data. 
This is not unexpected since in the small-$q_T$ region 
the convergence of the fixed-order
perturbative expansion is spoiled
by the presence 
of powers of large logarithmic terms, 
$\alpha_S^n\ln^m (m^2_V/q_T^2)$.
To obtain reliable predictions these terms have to be resummed to all orders.

The resummation is performed at the level of the partonic cross section, which is decomposed as
${d{\hat \sigma^V}_{ab}}/{dq_T^2}=
{d{\hat \sigma}^{V\,(\rm res.)}_{ab}}/{dq_T^2}
\,+\,{d{\hat \sigma}^{V\,(\rm fin.)}_{ab}}/{dq_T^2}\, .$
The term
$d{\hat \sigma}^{V\,({\rm res.})}_{ab}$
contains all the logarithmically enhanced contributions (at small $q_T$),
which have to be resummed
to all orders in $\alpha_S$,
while the term $d{\hat \sigma}^{V\,({\rm fin.})}_{ab}$
is free of such contributions and
can  be evaluated at fixed order in perturbation theory. 
Using the Bessel transformation between the conjugate variables 
$q_T$ and $b$,
the resummed component $d{\hat \sigma}^{V\,({\rm res.})}_{ab}$
can be expressed as
\begin{equation}
\label{resum}
\frac{d{\hat \sigma}_{ab}^{V\,(\rm res.)}}{dq_T^2}(q_T,M,\hat s,\alpha_S) 
=\hat\sigma^{V}_{LO}(M)\, \frac{M^2}{\hat s} \;
\int_0^\infty db \; \frac{b}{2} \;J_0(b q_T) 
\;{\cal W}_{ab}^{V}(b,M,\hat s,\alpha_S) \;,
\end{equation}
where $\hat\sigma^{V}_{LO}$ is the Born partonic cross section and
$J_0(x)$ is the $0$-order Bessel function.
By taking the $N$-moments of ${\cal W}$ with respect to the 
variable $z=M^2/{\hat s}$ at fixed $M$,
the resummation structure of ${\cal W}_{ab, \,N}^V$ can be 
organized in the exponential 
form~\footnote{For the sake of simplicity, here we consider only
the case of 
the diagonal terms in the flavour space. For the general case and a 
detailed discussion of the resummation formalism see
Ref.~\cite{hqt}.}
\begin{equation}
\label{wtilde}
{\cal W}_{N}^{V}(b,M,\alpha_S)
={\cal H}_{N}^{V}\left(\alpha_S \right) %\nonumber \\
\times \exp\{{\cal G}_{N}(\alpha_S,L)\}
\;,\;~ 
\mbox{with}\;~ L= \ln ({Q^2 b^2}/{b_0^2}),\;~
b_0=2e^{-\gamma_E}\,.
\end{equation}
The scale $Q\sim M \sim m_V$, that appears in the above formula
is named resummation scale 
and it parameterizes the
arbitrariness in the resummation procedure.
Variations of $Q$ around $m_V$ can be used to estimate the
size of higher-order
logarithmic contributions that are not explicitly resummed in a given 
calculation.

The process dependent function ${\cal H}_N^{V}$ 
includes all the perturbative
terms that behave as constants as $q_T\to 0$. %$b\to\infty$ (i.e. ). 
It can thus be expanded in powers of $\alpha_S=\alpha_S(\mu_R^2)$:
\begin{equation}
\label{hexpan}
{\cal H}_N^{V}(\alpha_S)=
\Bigl[ 1+ \frac{\alpha_S}{\pi} \,{\cal H}_N^{V \,(1)} 
\left(\frac{\alpha_S}{\pi}\right)^2 
\,{\cal H}_N^{V \,(2)}+\dots \Bigr]\,.
\end{equation}
The universal exponent ${\cal G}_N$ 
contains all
the terms that order-by-order in $\alpha_S$ are logarithmically divergent 
as $b\to\infty$ (i.e. $q_T\to 0$). 
The logarithmic expansion of ${\cal G}_N$ reads
\begin{equation}
\label{exponent}
{\cal G}_{N}(\alpha_S,L)=L g^{(1)}(\alpha_S L)+g_N^{(2)}(\alpha_S L)
+\frac{\alpha_S}{\pi} g_N^{(3)}(\alpha_S L)+\dots
\end{equation}
where the term $L\, g^{(1)}$ collects the leading logarithmic (LL) 
contributions, the function $g_N^{(2)}$ includes
the next-to-leading leading logarithmic
(NLL) contributions 
and so forth
~\footnote{To reduce the impact 
of unjustified higher-order contributions in the large-$q_T$ region,
the logarithmic variable $L$ in Eq.~(\ref{wtilde}), which diverges
for $b\to 0$, is replaced 
by $\tL\equiv \ln \left({Q^2 b^2}/{b_0^2}+1\right)$.
As a consequence of this replacement,
integrating the 
$q_T$  distribution over $q_T$
we obtain the corresponding total cross section:
$\int_0^\infty dq_T^2 ({d\hat \sigma}/{dq_T^2})_{\mbox{\itshape \tiny NLL+LO}}
 = \hat \sigma^{(tot)}_{\mbox{\itshape \tiny NLO}}$.}.

Finally 
the finite component 
has to be evaluated
starting from the usual fixed-order perturbative truncation 
of the partonic cross section
and subtracting the expansion of the resummed part at the same 
perturbative order: 
$[d{\hat \sigma^{V\,({\rm fin.})}_{ab}}/{dq_T^2}]_{f.o.}=
[d{\hat \sigma^{V}_{ab}}/{dq_T^2}]_{f.o.}
\,-\,
[d{\hat \sigma^{V\,({\rm res.})}_{ab}}/{dq_T^2}]_{f.o.}
$.
This matching procedure 
between resummed and finite contributions guarantees
to achieve uniform theoretical accuracy 
over the entire range
of transverse momenta.

The inclusion of the functions $g^{(1)}$, $g_N^{(2)}$,
${\cal H}_N^{V(1)}$ in the resummed component
and of the finite component at LO (i.e. ${\cal O}(\alpha_S)$)
allows us to perform the resummation at NLL+LO accuracy.
The inclusion of the functions $g_N^{(3)}$ and ${\cal H}_N^{V(2)}$ and of  
the finite component at NLO leads to a full NNLL+NLO accuracy.
Since the  coefficient ${\cal H}_N^{V(2)}$ has been computed only 
recently~\cite{wII}, here we limit ourselves to presenting results  
up to NLL+LO accuracy.

%%%%

\begin{figure}[h]
\begin{center}
\psfig{figure=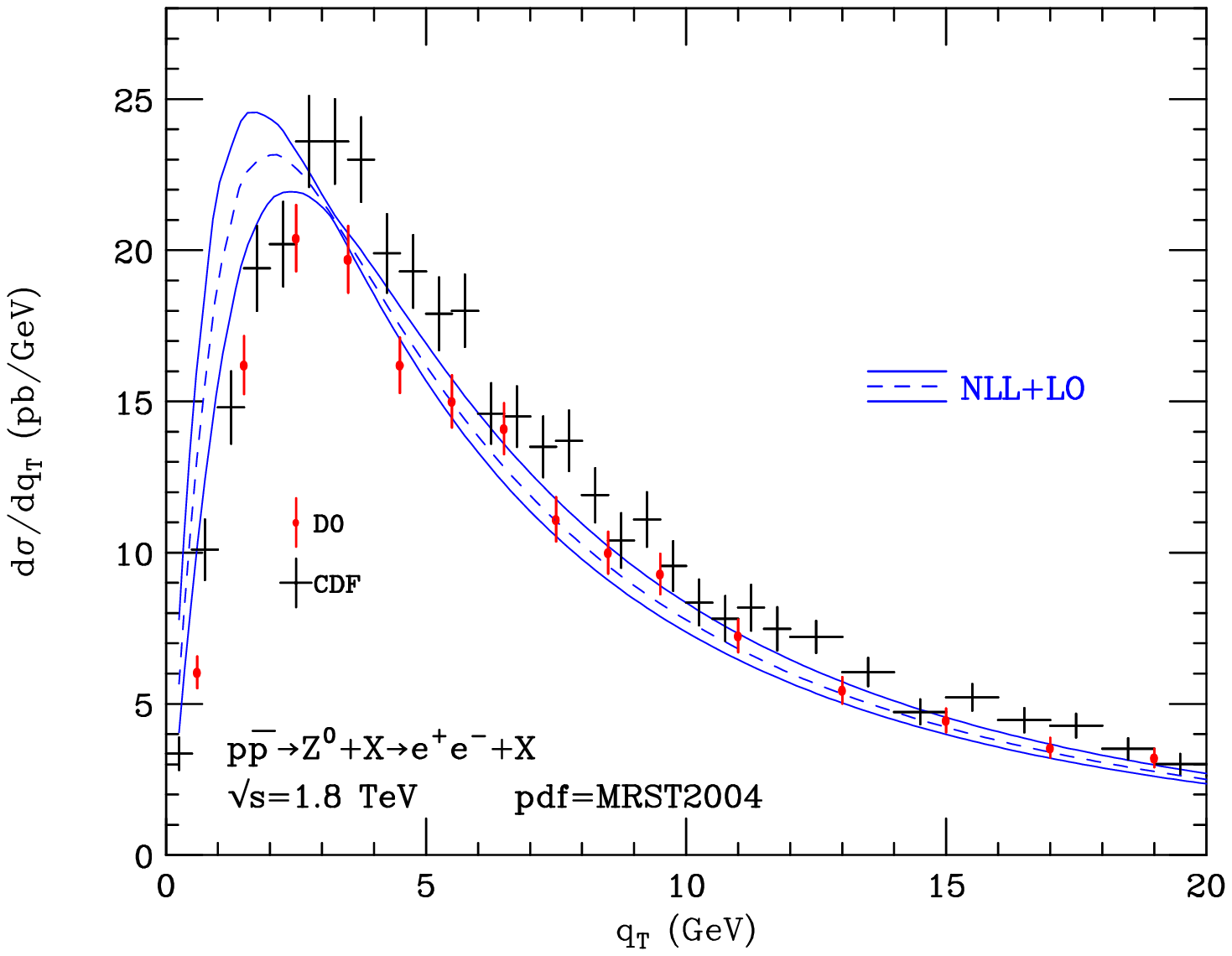,height=2.1in}
\hskip 1cm
\psfig{figure=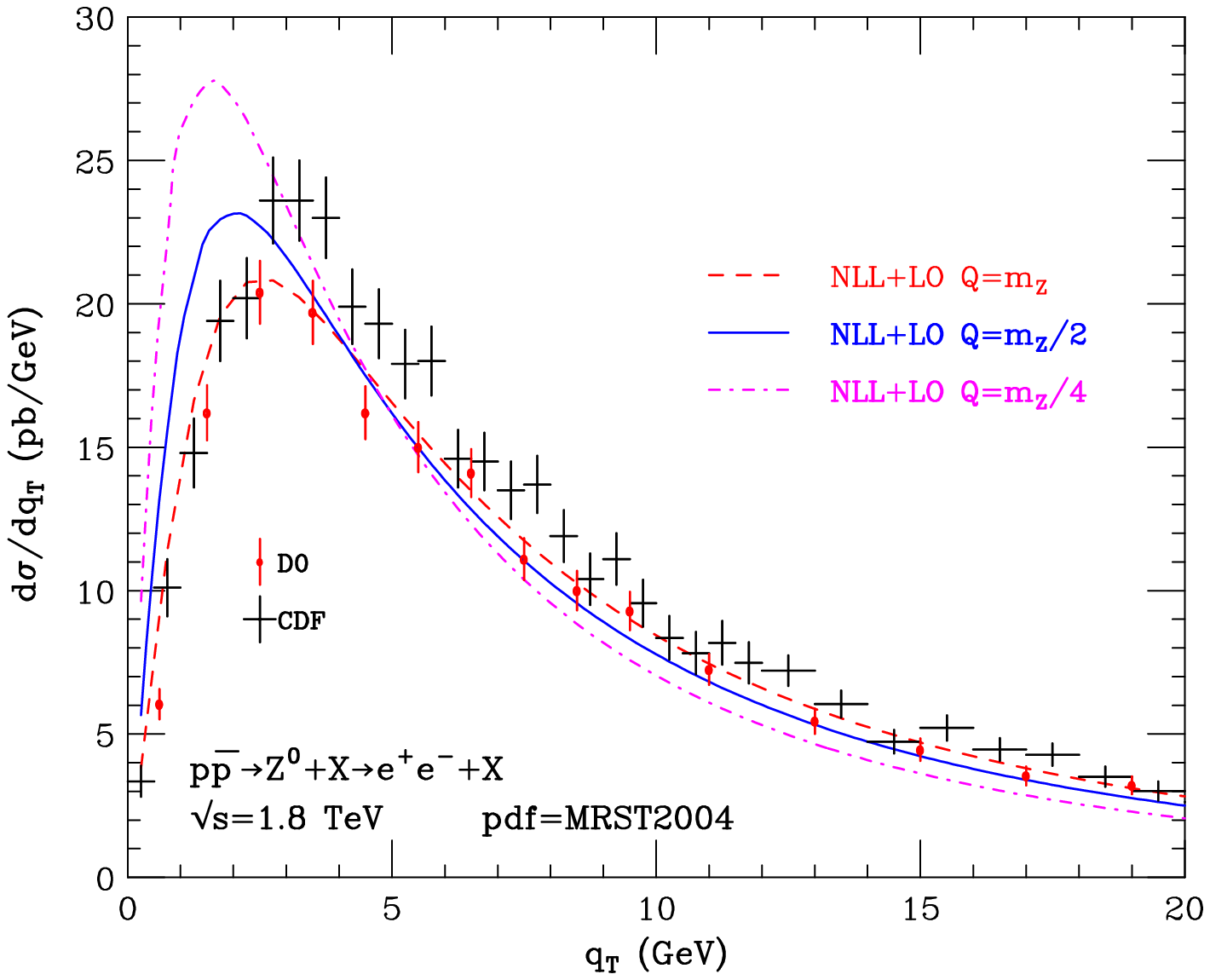,height=2.1in}
\label{fig2}%}
\vspace*{-.3cm}
\caption{
The $q_T$-spectrum of the Drell-Yan $e^+ e^-$ pairs produced in $p\bar p$ 
collisions at the Tevatron Run I. %~\cite{data}. 
Theoretical results are shown at NLL+LO, %(red dashed lines)
%and NLO (blue solid lines) 
including %the following 
scale 
variations. %
Left side: $m_Z/2\leq\mu_F , \mu_R \leq  2\;m_Z$ , with the
constraint $1/2 \leq \mu_F /\mu_R \leq 2$.
Right side:
$m_Z/4 \leq Q \leq m_Z$}
\end{center}
\end{figure}

In Fig.~\ref{fig2} we compare our NLL+LO resummed spectrum~\cite{wI} 
(with different values of the facto\-rization, renormalization 
and resummation scale) % $Q$.
with the Tevatron %the CDF and D0 Run I 
data.
We find that the scale
uncertainty is about $\pm 12-15\%$ from the region of the peak up to the
intermediate $q_T$ region ($q_T\sim 20$ GeV), and it is 
dominated by the resummation-scale uncertainty.
Taking into account the scale uncertainty,
we see 
that the resummed curve 
agrees reasonably well with the experimental points. 
We expect a sensible reduction of the scale dependence once the complete
NNLL+NLO calculation is available.

We note that in Fig.~\ref{fig2} the theoretical results are obtained in a
pure perturbative framework, without introducing any models of
non-perturbative contributions.
These contributions can be relevant in the $q_T$ region below the peak.
%
%\clearpage
\section{Fully exclusive NNLO Drell-Yan calculation}

We now consider QCD radiative corrections at the fully exclusive level for the
process in Eq.~\ref{first}. The purpose is to compute
observables $d{\hat\sigma}^{V}$, with arbitrary (though infrared safe)
kinematical cuts on the final-state leptons and the associated jet activity.
Provided the observable is sufficiently inclusive over the small-$q_T$ 
region, it can reliably be computed at fixed order in perturbation theory.

Following Ref.~\cite{hnnlo}, we observe that, at LO, 
the transverse momentum $q_T$ of $V$ is exactly zero.
This means that if $q_T\neq 0$
the (N)NLO contributions is given by the (N)LO contribution to the final state
$V + jet(s)$:
$\,
d{\hat\sigma}^{V}_{(N)NLO}|_{q_T\neq 0}=d{\hat\sigma}^{V+{\rm jets}}_{(N)LO}\,\,$.
We compute $d{\hat\sigma}^{V+{\rm jets}}_{NLO}$ by using the 
subtraction method at NLO and
we treat the remaining NNLO singularities at $q_T = 0$ by the additional 
subtraction of a counter-term%
~\footnote{The explicit
form of the counter-term $d{\hat\sigma}^{CT}_{(N)LO}$ 
is given in Ref.\cite{hnnlo}.} 
constructed by exploiting the universality of the
logarithmically-enhanced contributions to the $q_T$ distribution
(see Eq.~\ref{wtilde})
\begin{equation}
\label{main}
d{\hat\sigma}^{V}_{(N)NLO}=%\frac{1}{\hat\sigma^V_{LO}}
{\cal H}^{V}_{(N)NLO}\otimes d{\hat\sigma}^{V}_{LO}
+\left[ d{\hat\sigma}^{V+{\rm jets}}_{(N)LO}-
d{\hat\sigma}^{CT}_{(N)LO}\right]\;\; ,
\end{equation}
where $\mathcal{H}^V_{(N)NLO}$ is the process dependent coefficient function
of Eq.~\ref{hexpan}.

\begin{figure}[h]
\begin{center}
\psfig{figure=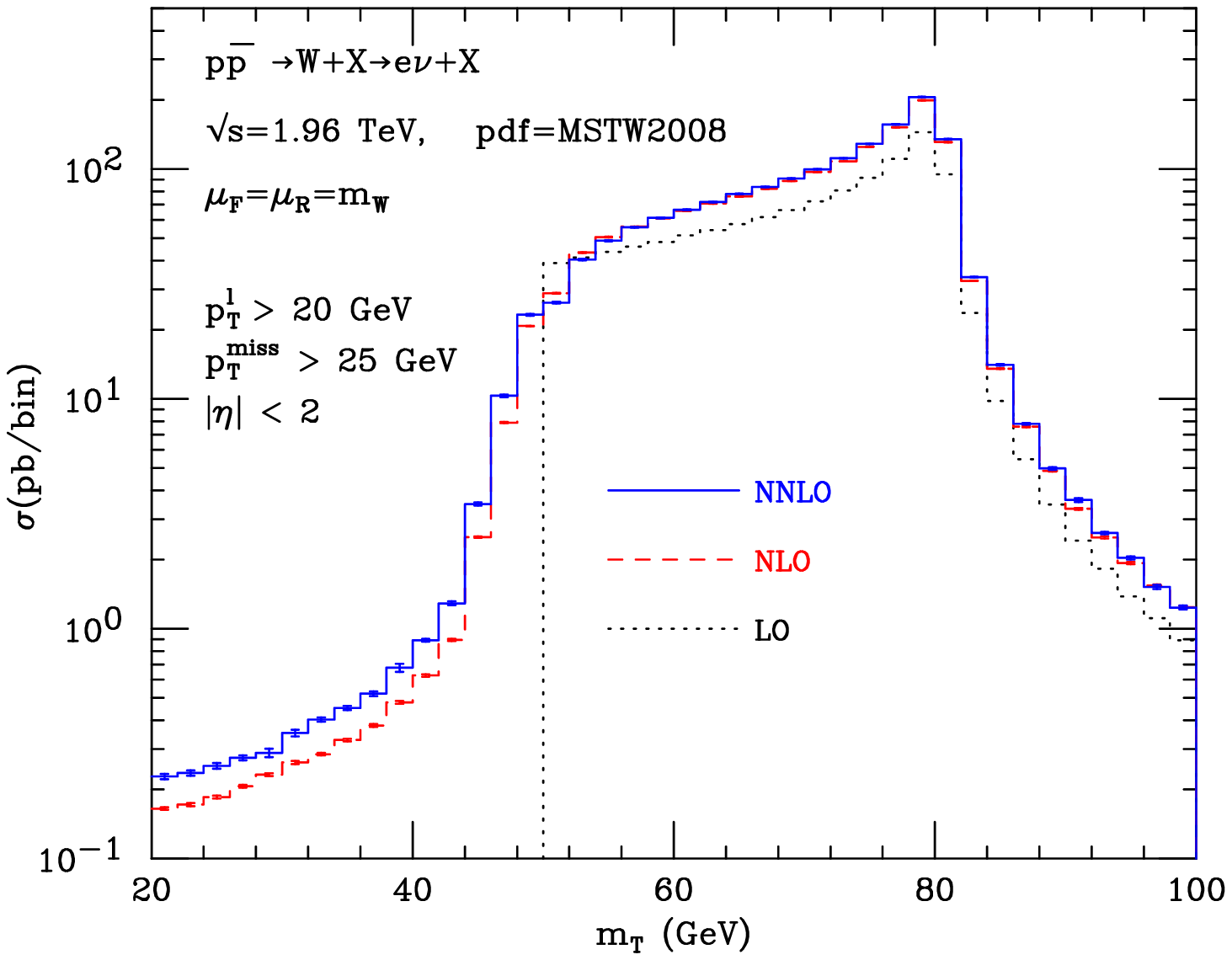,height=2.1in}
\hskip 1cm
\psfig{figure=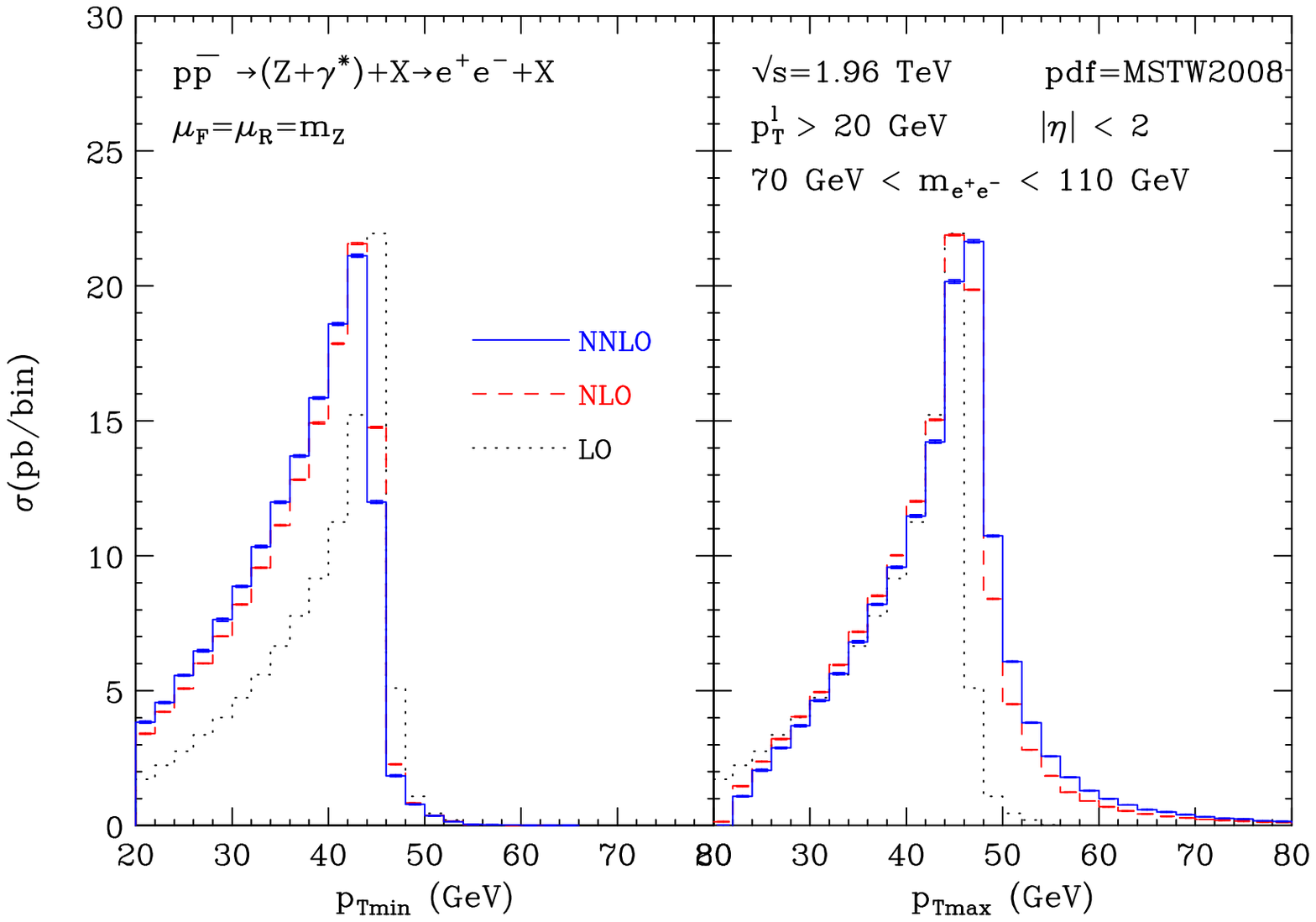,height=2.1in}
\caption{Left side: transverse mass distribution for $W$ production 
at the Tevatron.
Right side: distributions in $p_{T\,min}$ and $p_{T\,max}$ for the $Z$ signal 
at the Tevatron.
\label{figDYNNLO}}
\end{center}
\end{figure}

   We have encoded our NNLO computation in a parton level 
Monte Carlo event generator. 
The calculation includes finite-width effects,
the $\gamma-Z$ interference,
the leptonic decay of the vector bosons and the corresponding 
spin correlations. 
Our numerical code is particularly suitable
for the computation 
of distributions in the form of bin histograms, as shown the
illustrative  numerical results presented in Fig.~\ref{figDYNNLO}.


\vspace*{-.5cm}
\section*{References}
\begin{thebibliography}{99}
\bibitem{wI}
  G.~Bozzi, S.~Catani, G.~Ferrera, D.~de Florian and M.~Grazzini,
  %``Transverse-momentum resummation: a perturbative study of Z production at
  %the Tevatron,''
  Nucl.\ Phys.\  B {\bf 815} (2009) 174
  [arXiv:0812.2862 [hep-ph]].
  %%CITATION = NUPHA,B815,174;%%
\bibitem{wII}
  S.~Catani, L.~Cieri, G.~Ferrera, D.~de Florian and M.~Grazzini,
  %``Vector boson production at hadron colliders: a fully exclusive QCD
  %calculation at NNLO,''
  arXiv:0903.2120 [hep-ph].
  %%CITATION = ARXIV:0903.2120;%%
\bibitem{hnnlo}
S.~Catani and M.~Grazzini,
%``An NNLO subtraction formalism in hadron collisions and its application   to
%Higgs boson production at the LHC,''
Phys.\ Rev.\ Lett.\  {\bf 98} (2007) 222002.
\bibitem{nlo} %R.~K.~Ellis, G.~Martinelli and R.~Petronzio,
%``Lepton Pair Production At Large Transverse Momentum In Second Order QCD,''
%Nucl.\ Phys.\  B {\bf 211} (1983) 106;
%%CITATION = NUPHA,B211,106;%%
P.~B.~Arnold and M.~H.~Reno,
%``The Complete Computation of High p(t) W and Z Production in 2nd Order
%QCD,''
Nucl.\ Phys.\  B {\bf 319}, 37 (1989)
[Erratum-ibid.\  B {\bf 330}, 284 (1990)];
%%CITATION = NUPHA,B319,37;%%
R.~J.~Gonsalves, J.~Pawlowski and C.~F.~Wai,
%``QCD RADIATIVE CORRECTIONS TO ELECTROWEAK BOSON PRODUCTION AT LARGE
%TRANSVERSE MOMENTUM IN HADRON COLLISIONS,''
Phys.\ Rev.\  D {\bf 40}, 2245 (1989).
%%CITATION = PHRVA,D40,2245;%%

\bibitem{data}
A.~A.~Affolder {\it et al.}  [CDF Collaboration],
%``The transverse momentum and total cross section of $e^+e^-$ pairs in the
%$Z$ boson region from $p\bar{p}$ collisions at $\sqrt{s} = 1.8$ TeV,''
Phys.\ Rev.\ Lett.\  {\bf 84} (2000) 845;
%[arXiv:hep-ex/0001021].
%%CITATION = PRLTA,84,845;%%
%\bibitem{Abbott:1999tt}
B.~Abbott {\it et al.}  [D0 Collaboration],
%``Extraction of the width of the $W$ boson from measurements of
%$\sigma(p\bar{p} \to W + X) \times B(W \to e \nu)$ and $\sigma(p\bar{p} \to Z
%+ X) \times B(Z \to e e)$ and their ratio,''
Phys.\ Rev.\  D {\bf 61}, 072001 (2000).
%[arXiv:hep-ex/9906025].
%%CITATION = PHRVA,D61,072001;%%

\bibitem{hqt}
G.~Bozzi, S.~Catani, D.~de Florian and M.~Grazzini,
%``Transverse-momentum resummation and the spectrum of the Higgs boson at the
%LHC,''
Nucl.\ Phys.\ B {\bf 737} (2006) 73,
%[arXiv:hep-ph/0508068].
%%CITATION = HEP-PH 0508068;%%
%\cite{Bozzi:2003jy}
%G.~Bozzi, S.~Catani, D.~de Florian and M.~Grazzini,
%``The q(T) spectrum of the Higgs boson at the LHC in QCD perturbation
%theory,''
Phys.\ Lett.\ B {\bf 564} (2003) 65,
%[arXiv:hep-ph/0302104].
%%CITATION = HEP-PH 0302104;%%
%\cite{Bozzi:2007pn}
%\bibitem{Bozzi:2007pn}
%G.~Bozzi, S.~Catani, D.~de Florian and M.~Grazzini,
%``Higgs boson production at the LHC: transverse-momentum resummation and
%rapidity dependence,''
Nucl.\ Phys.\  B {\bf 791} (2008) 1.
%[arXiv:0705.3887 [hep-ph]].
%%CITATION = NUPHA,B791,1;%%
%\bibitem{ja}C Jarlskog in {\em CP Violation}, ed. C Jarlskog
%(World Scientific, Singapore, 1988).
%\bibitem{ma}L. Maiani, \Journal{\PLB}{62}{183}{1976}.
%\bibitem{bu}J.D. Bjorken and I. Dunietz, \Journal{\PRD}{36}{2109}{1987}.
%\bibitem{bd}C.D. Buchanan {\it et al}, \Journal{\PRD}{45}{4088}{1992}.
%\cite{Catani:2007vq}
%\bibitem{hqt}
%G.~Bozzi, S.~Catani, D.~de Florian and M.~Grazzini,
\end{thebibliography}
\end{document}